\documentclass{moriond}
\usepackage{subfig}
\usepackage{amssymb,amsmath}
\usepackage{tikz}

\newlength\imageheight

\newcommand{\Photo}{}
\newcommand\pubnumber{PSI-PR-21-09, ZU-TH-20/21}
\newcommand\pubblock{\rightline{\begin{tabular}{l} \pubnumber
\end{tabular}}}

\begin{document}
\title{EXPLAINING THE CABIBBO ANGLE ANOMALY AND LEPTON FLAVOUR UNIVERSALITY VIOLATION IN TAU DECAYS WITH A SINGLY-CHARGED SCALAR SINGLET}
\author{Fiona Kirk}

\address{
Physik-Institut, Universit\"at Z\"urich, Winterthurerstrasse 190, CH--8057 Z\"urich, Switzerland\\
Paul Scherrer Institut, CH--5232 Villigen PSI, Switzerland}

\pubblock
\maketitle

\abstracts{
The singly charged $SU(2)_L$ singlet scalar, with its necessarily flavour violating couplings to leptons, lends itself particularly well for an explanation of the Cabibbo Angle Anomaly and of hints for lepton flavour universality violation in $\tau \to \mu\bar \nu\nu$. In a setup addressing both anomalies, we predict loop-induced effects in $\tau\to e\gamma$ and in $\tau \to e\mu\mu$. A recast of ATLAS selectron and smuon searches allows us to derive a coupling-independent lower limit of $\approx 200$ GeV on the mass of the singly charged singlet scalar. At a future $e^+e^-$ collider, dark matter mono-photon searches could provide a complementary set of bounds.
}

\section{Introduction}\label{Intro}
\begin{figure}[t]%
\settoheight{\imageheight}{\includegraphics[scale=.7]{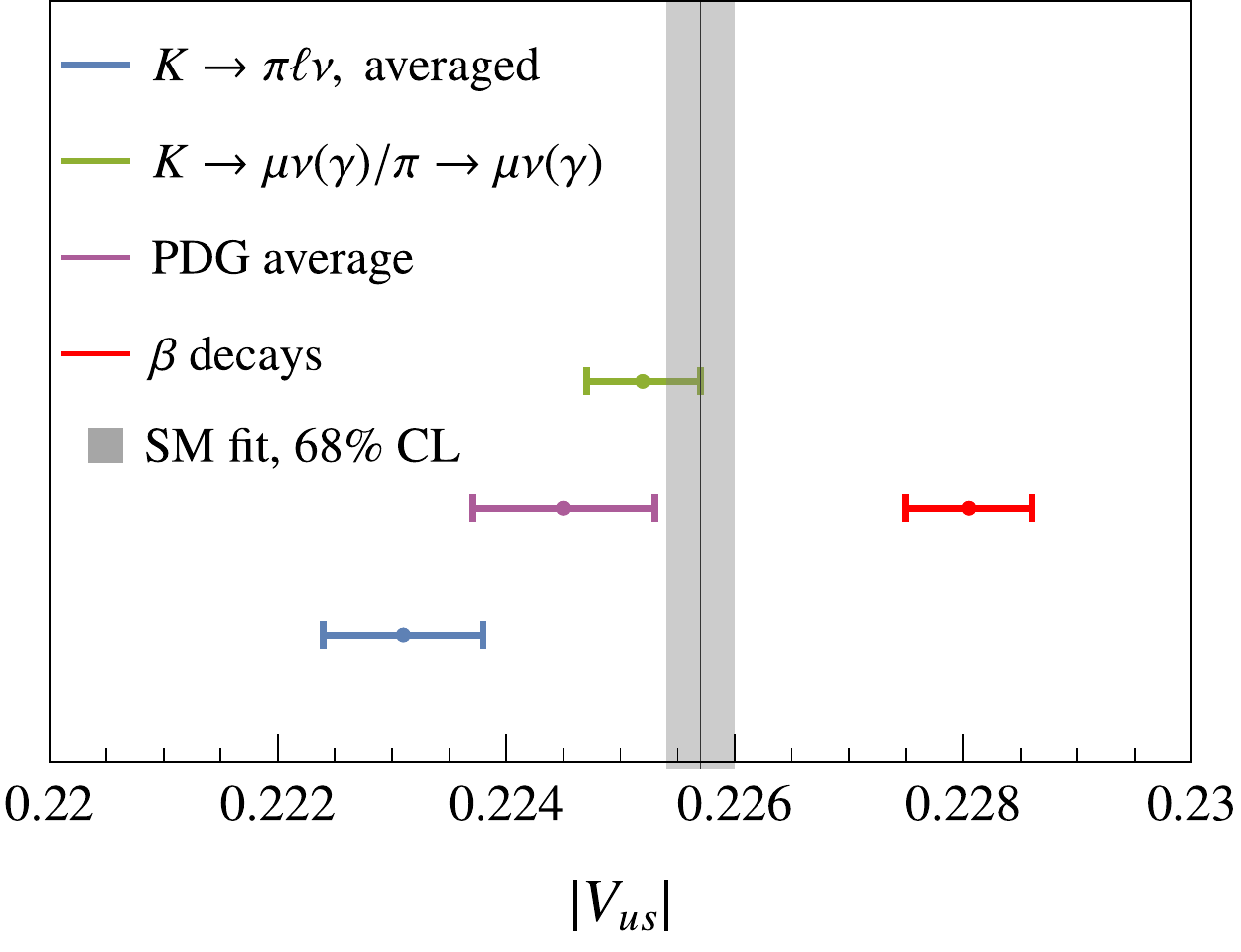}}
    \centering
\subfloat[\label{fig:CAAdata}]{\includegraphics[scale=.65]{CAA_data_2}}%
\hfil
\subfloat[\label{fig:CAANP}]{\tikz\node[minimum height=\imageheight]{\includegraphics[scale=.5]{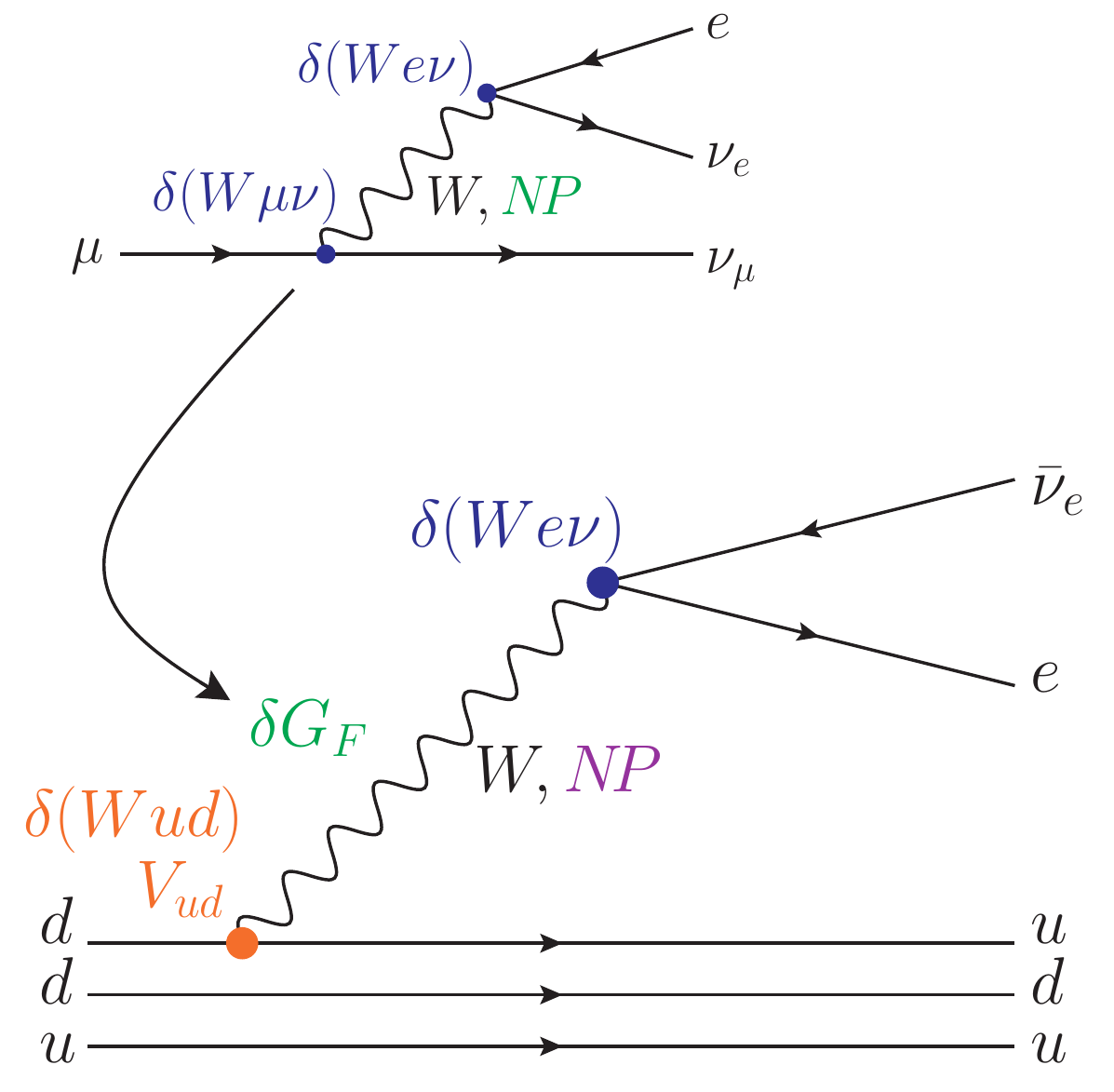}}; }%
\caption{(a) Determinations of $|V_{us}|$ from different sources.\protect\cite{Antonelli:2010yf,Aoki:2019cca,Zyla:2020zbs} 
$|V_{us}|=0.22805(64)\equiv |V_{us}^\beta|$, which was determined, using CKM unitarity from $|V_{ud}|=0.97365(15)$,\,\protect\cite{Seng:2020wjq} is shown in red.\\
 (b) Possible NP interpretations leading to effects in $\beta$-decays (and $\mu\to e\bar \nu\nu$). Note that the contribution of $\delta(We\nu)$ to $\mu\to e\bar \nu\nu$ cancels against the direct contribution of $\delta(We\nu)$ to the $\beta$-decay. $\delta(W\mu\nu)$, however, leads to an effect in $\beta$-decays.}
\label{fig:CAA}%
    \end{figure}

The Cabibbo angle, which describes the mixing of the first two generations of quarks in the Standard Model (SM), 
can be determined from $V_{ud}$, extracted from superallowed $\beta$-decays, from $V_{us}$ from $\tau$-decays or $K$-decays, or from $V_{cd}$ from $D\to \mu \nu$.
Recent improvements in the computation of the $\gamma W$-box contributions to superallowed $\beta$ decays reduced the uncertainty on $V_{ud}$,\cite{Seng:2018yzq} however, they also shifted the central value down to $|V_{ud}|=0.97370(14)$,\cite{PDGreview} suggesting a 3$\sigma$ deviation from first-row Cabibbo Kobayashi Maskawa (CKM) unitarity.\cite{Seng:2018yzq,Czarnecki:2019iwz,Hayen:2020nej,Shiells:2020fqp,Zyla:2020zbs} The tension between $|V_{us}|$ extracted from $K$- or $\tau$-decays, and $|V_{us}|$ determined using (first row) CKM unitarity ($|V_{ud}|^2+|V_{us}|^2+|V_{ub}|^2=1$) from $|V_{ud}|$, which is extracted from superallowed $\beta$ decays, is commonly referred to as the Cabibbo Angle Anomaly (CAA)\cite{Belfatto:2019swo,tan2020laboratory,Grossman:2019bzp,Crivellin:2020lzu,Crivellin:2021njn}. In Fig.~\ref{fig:CAAdata} we show the values for $|V_{us}|$ quoted by the PDG,\cite{Antonelli:2010yf,Aoki:2019cca,Zyla:2020zbs} as well as the value $|V_{us}|=0.22805(64)\equiv |V_{us}^\beta|$, which was determined, using (first row) CKM unitarity ($|V_{ud}|^2+|V_{us}|^2+|V_{ub}|^2=1$) from $|V_{ud}|=0.97365(15)$.\,\cite{Seng:2020wjq}

\begin{figure}[t]
\settoheight{\imageheight}{\includegraphics[scale=1.3]{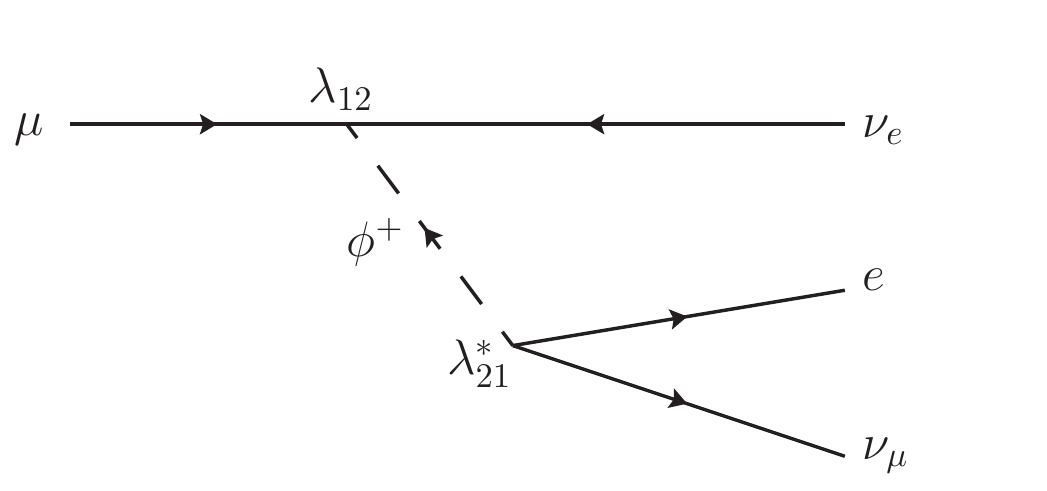}}
\centering
\subfloat[\label{fig:LFUtau}]{%
\includegraphics[scale=.55]{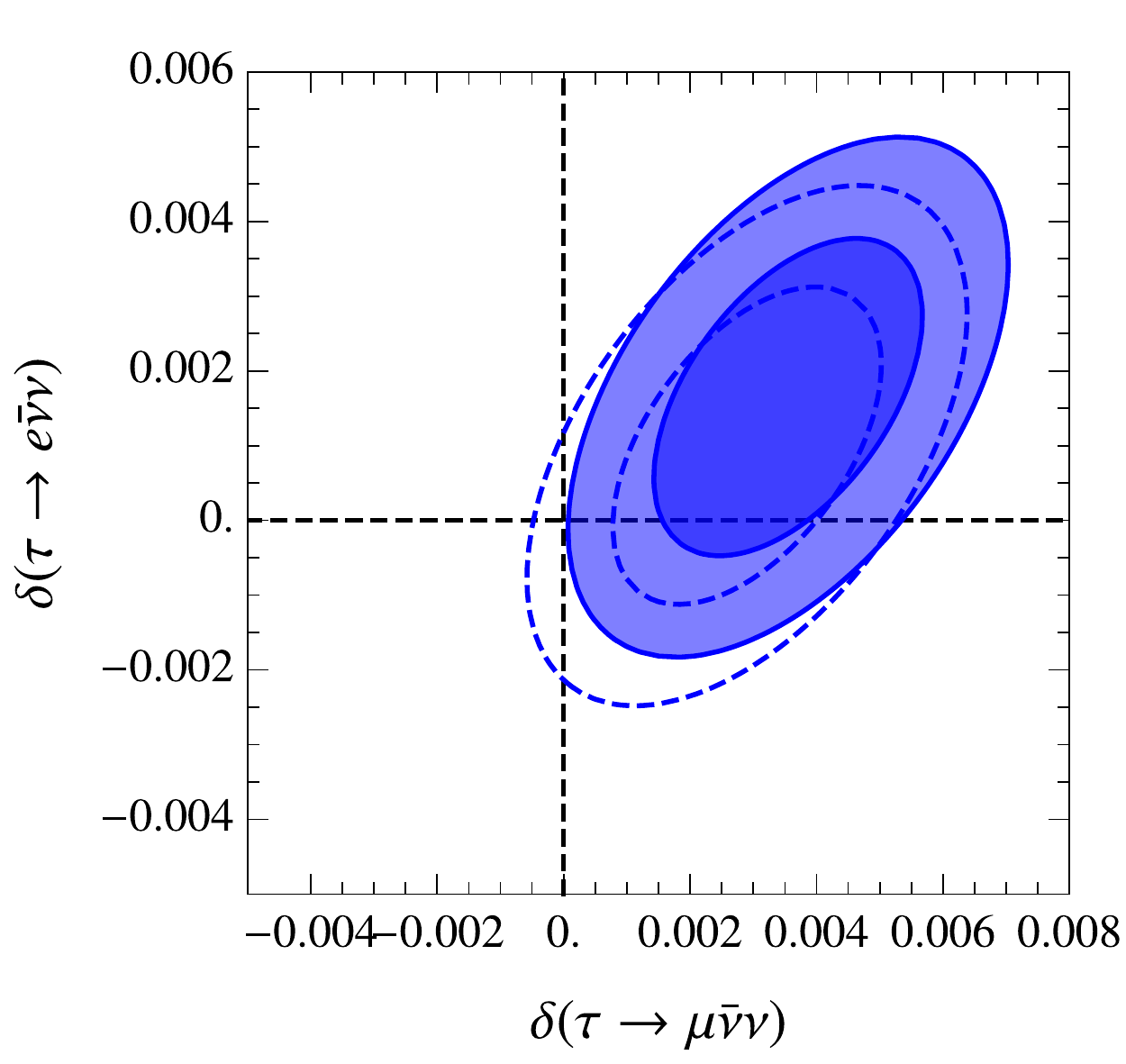}
}
\hspace{0.02\textwidth}
\subfloat[$\mu \to e \bar \nu\nu$\label{fig:llnunu}]{%
	{\tikz\node[minimum height=\imageheight]{
	\includegraphics[width=0.45\textwidth]{l_lnunu}};}%
}\caption{(a) Regions preferred by the ratios in E.q.~\eqref{eq:LFUratios} in terms of $\delta(\tau\to\mu\bar \nu\nu)$ and $\delta(\tau\to e\bar \nu\nu)$. The dashed lines show the 1$\sigma$ and 2$\sigma$ contours for $\delta(\mu\to e\bar \nu\nu)=0$, the blue regions show the 1$\sigma$ and 2$\sigma$ preferred regions for $\delta(\mu\to e\bar \nu\nu)=0.00065$, which is the value favoured by our global fit.\protect\cite{Crivellin:2020klg,deBlas:2019okz}
(b) Feynman diagram showing the contribution of $\phi^\pm$ to muon decay. (The corresponding diagram for tau decays is found by exchanging the flavour indices.)
}
\end{figure}

A number of models involving physics beyond the SM can account for this experimental situation,\cite{Kirk:2020wdk} however, they all adopt one of the following strategies\cite{Crivellin:2021njn} (see also Fig.~\ref{fig:CAANP} for an illustration): they
\begin{itemize}
\item 
lead to direct contributions to $\beta$-decays and thus affect the extraction of $V_{us}$ from $\beta$-decays\,\cite{Crivellin:2021rbf,Crivellin:2021egp} (shown in purple in Fig.~\ref{fig:CAANP}),
\item 
modify the $W\!ud$-coupling, leading to a violation of CKM unitarity\,\cite{Belfatto:2019swo,Belfatto:2021jhf,Branco:2021vhs} (orange),
\item 
modify $W\!\ell\nu$-couplings, which enter $\beta$-, $K$-, $\pi$- and $\mu$-decays\,\cite{Coutinho:2019aiy,Bryman:2019bjg,Crivellin:2020lzu,Crivellin:2020ebi,Capdevila:2020rrl,Endo:2020tkb,Alok:2020jod} (blue\,\footnote{Note that in order to have an effect in $\beta$-decays, we need a modification of the $W\mu\nu$-coupling, since the modification of the $We\nu$-coupling drops out due to the concurrent modification of the Fermi constant.}), or
\item 
lead to new contributions to $\mu\to e\bar\nu\nu$, which modify the Fermi constant and enter the extraction of $V_{ud}$ from superallowed beta decays in that way\cite{Belfatto:2019swo,Crivellin:2020klg,Crivellin:2020oup,Marzocca:2021azj,Buras:2021btx}(green).
\end{itemize}
Here we consider the last of these possibilities.\footnote{Indeed, the singly charged scalar can generate a new contribution to $\mu\to e\bar \nu\nu$ (see Fig.~\ref{fig:llnunu}).} Denoting the new physics (NP) $\mu\to e\bar \nu\nu$ amplitude, normalised to the SM amplitude, by
\begin{equation}
\delta(\mu\to e\bar \nu\nu)=\frac{\mathcal{A}_{NP}(\mu\to e\bar \nu\nu)}{\mathcal{A}_{SM}(\mu\to e\bar \nu\nu)},\label{deltamu}
\end{equation}
the Fermi constant, $G_F$, which is extracted from $\mu\to e\bar\nu\nu$, takes the form
\begin{equation}
G_F=G_F^{\rm SM}(1+\delta(\mu\to e\bar \nu\nu))\,.
\end{equation}
where $G_F^{\rm SM}$ stands for the Fermi constant determined in presence of SM physics only.
A modification of the Fermi constant can alleviate the tension between the different determinations of $V_{us}$ by shifting $V_{ud}^\beta$, i.e. $V_{ud}$ extracted from $\beta$-decays, from $V_{ud}^\beta=V_{ud}^{\rm SM}$, where $V_{ud}^{\rm SM}$ is the $(1,1)$-element of the (unitary) CKM matrix, under assumption of the SM,
to $V_{ud}^\beta=V_{ud}^{\rm SM}(1-\delta(\mu\to e \bar \nu\nu))$. Via CKM-unitarity this leads to
\begin{align}
V_{us}^\beta  &\equiv \sqrt {1 \!-\! {{\left|V_{ud}^\beta \right|}^2} \!-\! |{V_{ub}}{|^2}}  \simeq V_{us}^{\rm{SM}}\!\left[ {1 + {{\left( {\frac{{V_{ud}^{\rm{SM}}}}{{V_{us}^{\rm{SM}}}}} \right)}^2}\,\delta (\mu\to e\bar \nu\nu)} \right],
\label{Vusbeta}
\end{align}
which indeed is larger than $V_{us}^{\rm SM}$, if $\delta(\mu\to e\bar\nu\nu)$ is positive. Note that the enhancement by $\left(V_{ud}^{\rm{SM}}/V_{us}^{\rm{SM}}\right)^2\approx 19$ makes $V_{us}$ particularly sensitive to modifications of the Fermi constant from muon decay.\cite{Crivellin:2020lzu}

Besides the CAA, the singly charged scalar can address the 2$\sigma$ deviation from SM predictions encoded in the amplitude fractions~\cite{Amhis:2019ckw}
\begin{align}
\frac{{{\cal A}(\tau  \to \mu  \bar \nu \nu )}}{{{\cal A}(\mu  \to e \bar \nu\nu )}} = 1.0029(14){\mkern 1mu} ,\quad
\frac{{{\cal A}(\tau  \to \mu  \bar \nu\nu )}}{{{\cal A}(\tau  \to e \bar\nu \nu )}}  = 1.0018(14){\mkern 1mu} ,\quad
\frac{{{\cal A}(\tau  \to e \bar \nu \nu )}}{{{\cal A}(\mu  \to e\bar \nu \nu )}}  = 1.0010(14){\mkern 1mu}.
\label{eq:LFUratios}
\end{align}
Expressing this data in terms of $\delta(\tau\to \mu\bar \nu\nu)$ and $\delta(\tau\to e\bar \nu\nu)$, which are defined in analogy with $\delta(\mu\to e\bar \nu\nu)$ (see E.q.~\eqref{deltamu}),
we observe a preference for positive
$\delta(\tau \to \mu\bar \nu\nu)$ and $\delta(\tau \to e\bar \nu\nu)$ (see Fig.~\ref{fig:LFUtau}).

\section{{\boldmath The Singly Charged $SU(2)_L$ Singlet Scalar}}\label{SCS}
Let us now introduce the singly charged $SU(2)_L$ singlet scalar $\phi^+$, which can lead to the desired effects in muon and tau decays. This field has been extensively studied in the context of Zee models,\cite{Zee:1980ai,Wolfenstein:1980sy,Petcov:1982en,Zee:1985id} Babu-Zee models,\cite{Zee:1985id,Babu:1988ki,Chang:1988aa,Irie:2021obo} and other models where singly charged scalars radiatively generate neutrino masses,\cite{Balaji:2001ex,Dicus:2001ph,Drees:1997id,Bhattacharyya:1999tv,Cheung:1999az,Haba:1999iw,Cai:2017jrq,Krauss:2002px} however, in the following we will remain agnostic about the underlying theory and consider the singly charged scalar as a minimal extension of the SM.

The singly charged scalar is a $(1,1,1)$-representation of the SM gauge group $SU(3)_c\times SU(2)_L \times U(1)_Y$, which leaves only one way for  it to couple to SM matter fields:
\begin{align}
\mathcal{L}_{int} = - \frac{\lambda_{ij}}{2}\, \bar{L}^c_{a,i}\, \varepsilon_{ab}\, L_{b,j} \, \phi^+ + {\rm h.c.}\,.
\label{Lag}
\end{align}
Here $a,b$ are $SU(2)_L$ indices, $\varepsilon_{ab}$ is the 2-dimensional
Levi-Civita tensor, $c$ stands for charge conjugation,  $i,j$ are flavour
indices, and $\lambda_{ij}$ can be chosen to be antisymmetric in flavour without loss of generality. Hence this extension of the SM automatically leads to lepton flavour (universality) violation. The fact that only three new couplings and one mass are needed to fully describe the singly charged scalar, makes this model very predictive.

\subsection{Flavour Bounds}\label{Flavour}
\begin{figure*}
\subfloat[$\mu\to e \gamma$]{%
	\includegraphics[width=0.38\textwidth]{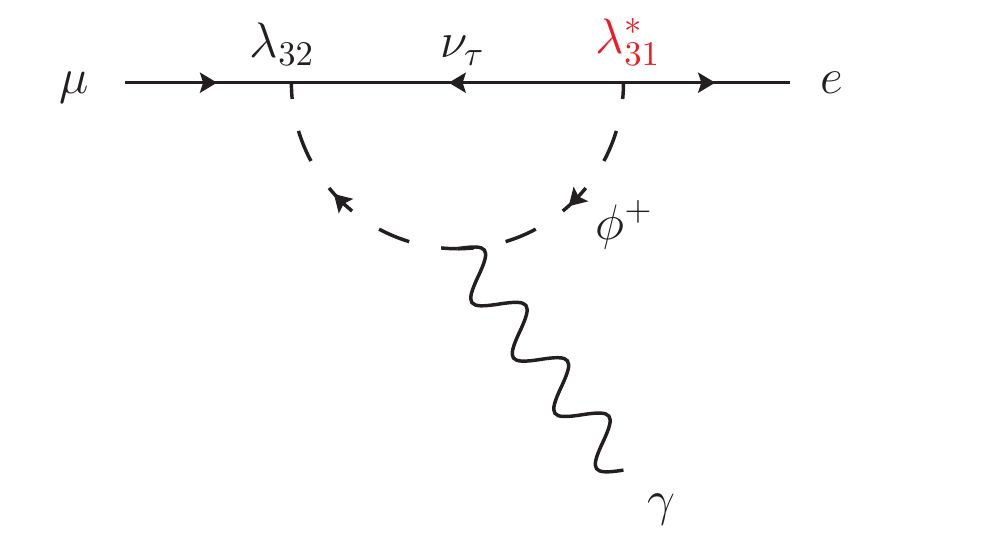}
}
\hspace{-0.04\textwidth}
\subfloat[$\tau \to e\bar\mu\mu$, I\label{fig:box1}]{%
	\includegraphics[width=0.3\textwidth]{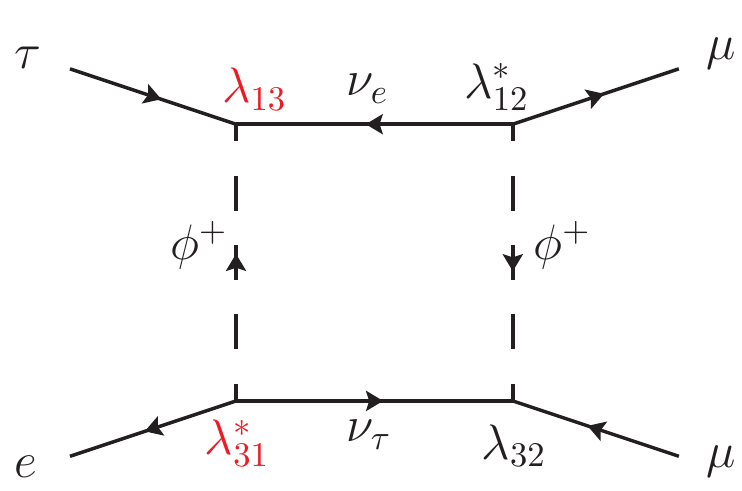}
}
\hspace{0\textwidth}
\subfloat[$\tau \to e\bar\mu\mu$, II\label{fig:box2}]{%
	\includegraphics[width=0.32\textwidth]{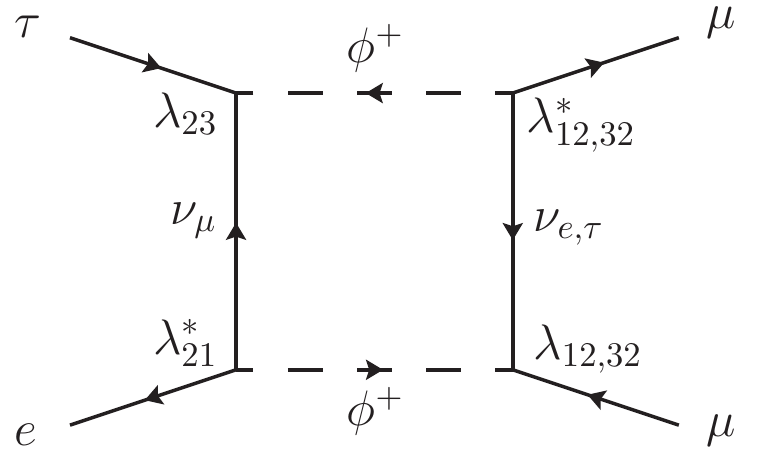}
	}
\caption{Feynman diagrams showing the contribution of the singly charged scalar $\phi^\pm$ to (a) radiative leptonic decays and (b,c) charged lepton flavour violation.
  The coupling $\lambda_{13}$, which is strongly constrained by $\mu \to e$ conversion and $\mu \to e\gamma$, is shown in red.
}\label{FeynmanDiagrams}
\label{fig:flavour}
\end{figure*}

Equipped with the Lagrangian in E.q.~\eqref{Lag}, we can derive the leading contributions of the singly charged scalar to the relevant flavour observables (see Figs.~\ref{fig:llnunu} and \ref{fig:flavour} for the corresponding Feynman diagrams). The first interesting observation we can make is that the modifications (see E.q.~\eqref{deltamu}) of the $\mu\to e\bar \nu\nu$, $\tau\to e\bar \nu\nu$ and $\tau\to \mu\bar \nu\nu$ amplitudes,
 \begin{align}
\delta(\ell_i\to \ell_j\bar \nu\nu)=\frac{\mathcal{A}_{NP}(\ell_i\to\ell_j\nu_i\bar\nu_j)}{\mathcal{A}_{SM}(\ell_i\to\ell_j\nu_i\bar\nu_j)} =\frac{\left\vert \lambda_{ij}\right\vert^2}{g_2^2}\frac{m_W^2}{m_\phi^2}\,,
\end{align}
are necessarily constructive. This fits to the CAA, in particular to E.q.~\eqref{Vusbeta}, and to the experimental data shown in E.q.~\eqref{eq:LFUratios} and in Fig~\ref{fig:LFUtau}. A global fit to EW data and the CKM elements states a preference for $\delta (\mu\to e\nu\nu)=0.00065(15)$.\cite{Crivellin:2020klg,deBlas:2019okz}

Radiative leptonic decays provide strong bounds on the couplings $\lambda_{ij}$ of the singly charged scalar to the SM leptons. In fact, the bound from $\mu\to e\gamma$ is so strong that we can set $\lambda_{13}\approx0$ (marked in red in Fig.~\ref{fig:flavour}). We reach a similar conclusion considering $\mu \to e $ conversion in nuclei. This has consequences for charged lepton violation, which is loop-suppressed in this model (see e.g. Figs.~\ref{fig:box1} and \ref{fig:box2}). Processes mediated by $\lambda_{13}$, such as $\tau\to e \bar \mu e$, are thus expected to be vanishingly small.

\enlargethispage{\baselineskip} 
\subsection{Collider Constraints}\label{Collider}
\begin{figure}
\settoheight{\imageheight}{\includegraphics[width=0.5\textwidth]{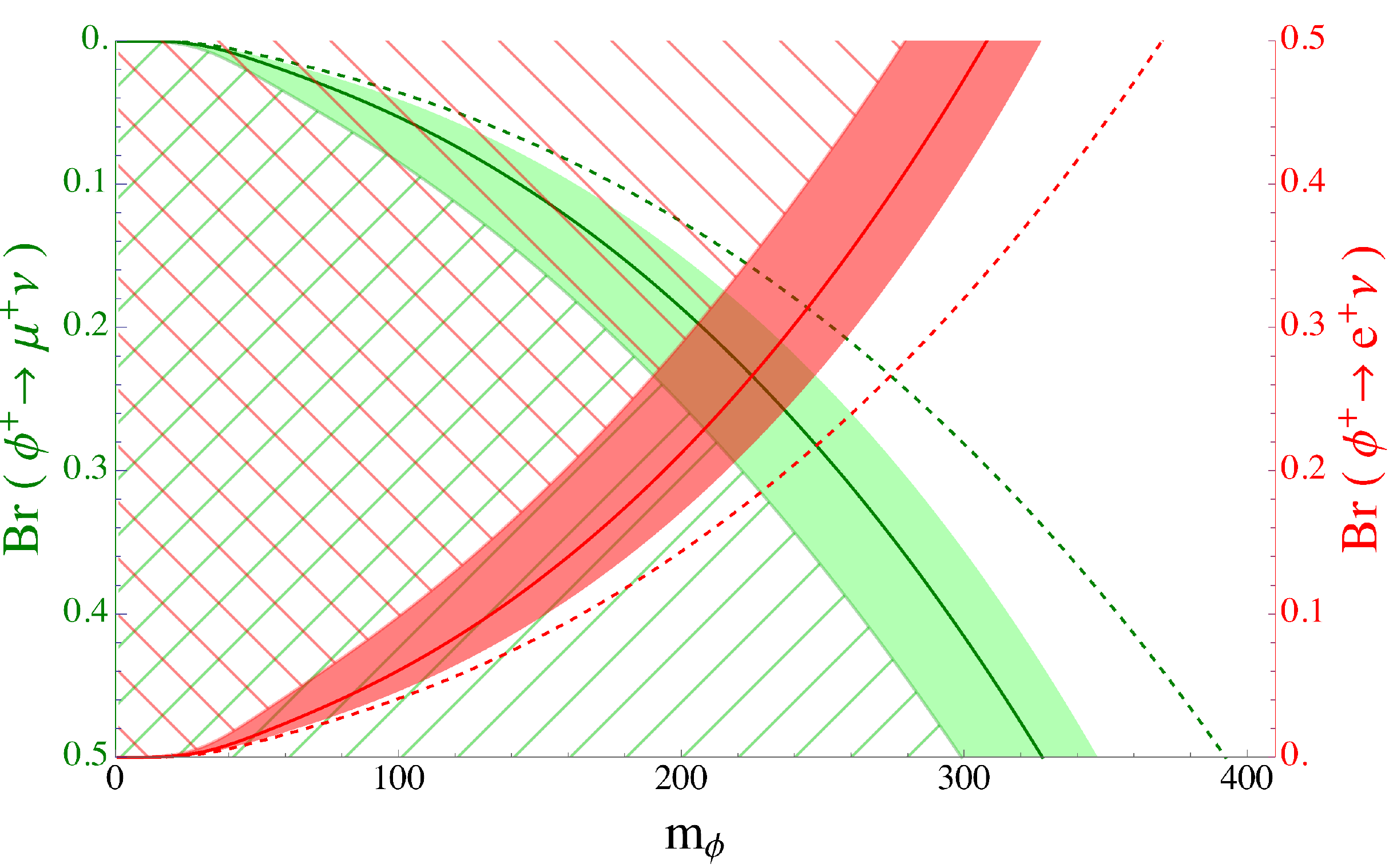}}
\centering
\subfloat[\label{fig:DrellYan}]{%
{\tikz\node[minimum height=\imageheight]{\includegraphics[width=0.4\textwidth]{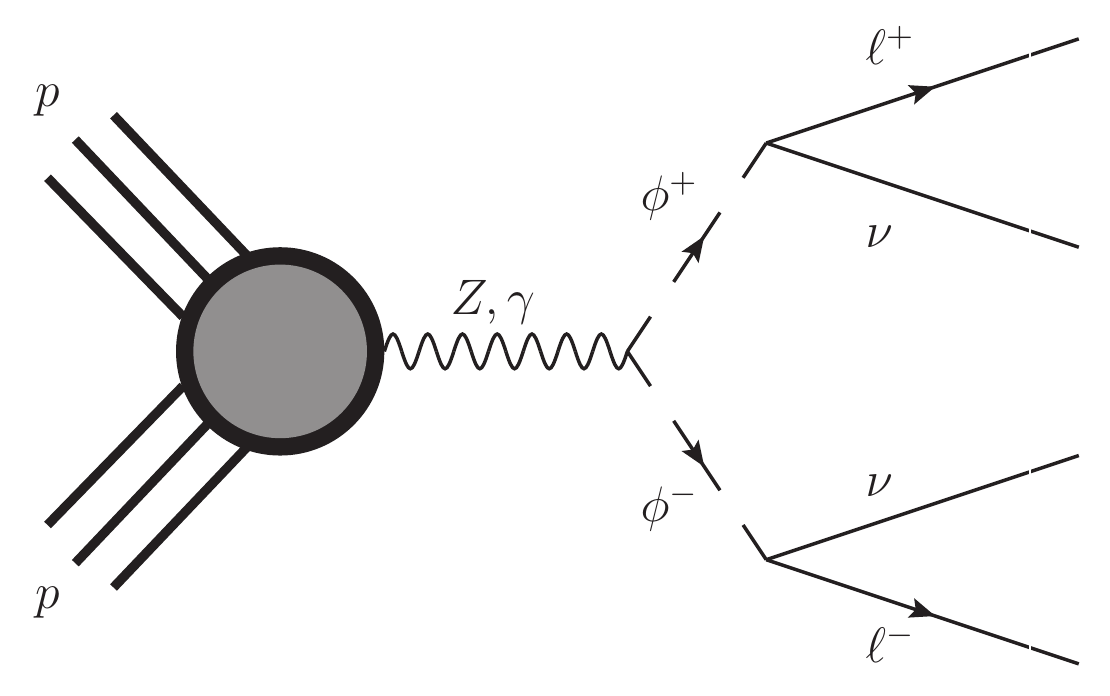}} ;}	
}
\hspace{0.02\textwidth}
\subfloat[\label{fig:ATLAS}]{%
	\includegraphics[width=0.52\textwidth]{LHC_BRsvsM_2} 
	}
\caption{(a) Feynman diagram showing Drell-Yan pair production of $\phi^\pm$ and their decay into SM charged leptons and SM neutrinos (leading to missing transverse energy in the detector)\\ (b) Recast ATLAS bounds (see text for the details)}	
\end{figure}

Since the singly charged scalar has the same quantum numbers as right-handed sleptons, we can simply recast selectron and smuon searches into searches for the singly charged scalar. 
The dominant channel here is Drell-Yan pair production of the singly charged scalar, which consecutively decays into a pair of charged leptons with missing transverse energy (see Fig~\ref{fig:DrellYan} for the corresponding Feynman diagram). 
Reinterpreting the most recent dedicated ATLAS search, which is based on 139~fb$^{-1}$ of proton-proton collisions,\,\cite{Aad:2019vnb} we obtain the bounds, shown in Fig.~\ref{fig:ATLAS}, on the mass $m_\phi$ of the singly charged scalar and on the branching ratios ${\rm Br}(\phi^+\to \mu^+ \nu)$ and ${\rm Br}(\phi^+\to e^+ \nu)$. 
The hatched regions are excluded by the $e^+e^-$ (red) or $\mu^+\mu^-$ (green) channels. The bounds depicted in Fig.~\ref{fig:ATLAS} allow us to set a coupling-independent lower limit of $m_\phi\approx 220$~GeV on the mass of the singly charged scalar. The scenario of $\lambda_{13}=0$, mentioned in Section~\ref{Flavour}, translates into a branching ratio of ${\rm Br}(\phi^+\to \mu^+\nu)=1/2$.
(The corresponding branching ratio into electrons cannot be fixed in this scenario without making additional assumptions on the relation between $\lambda_{12}$ and $\lambda_{23}$.) 
The projected exclusion limits of the High Luminosity (HL) LHC are indicated by dashed curves.

\section{Combined Analysis and Conclusions}\label{Combined}
\begin{figure*}
\centering
	\includegraphics[width=0.57\textwidth]{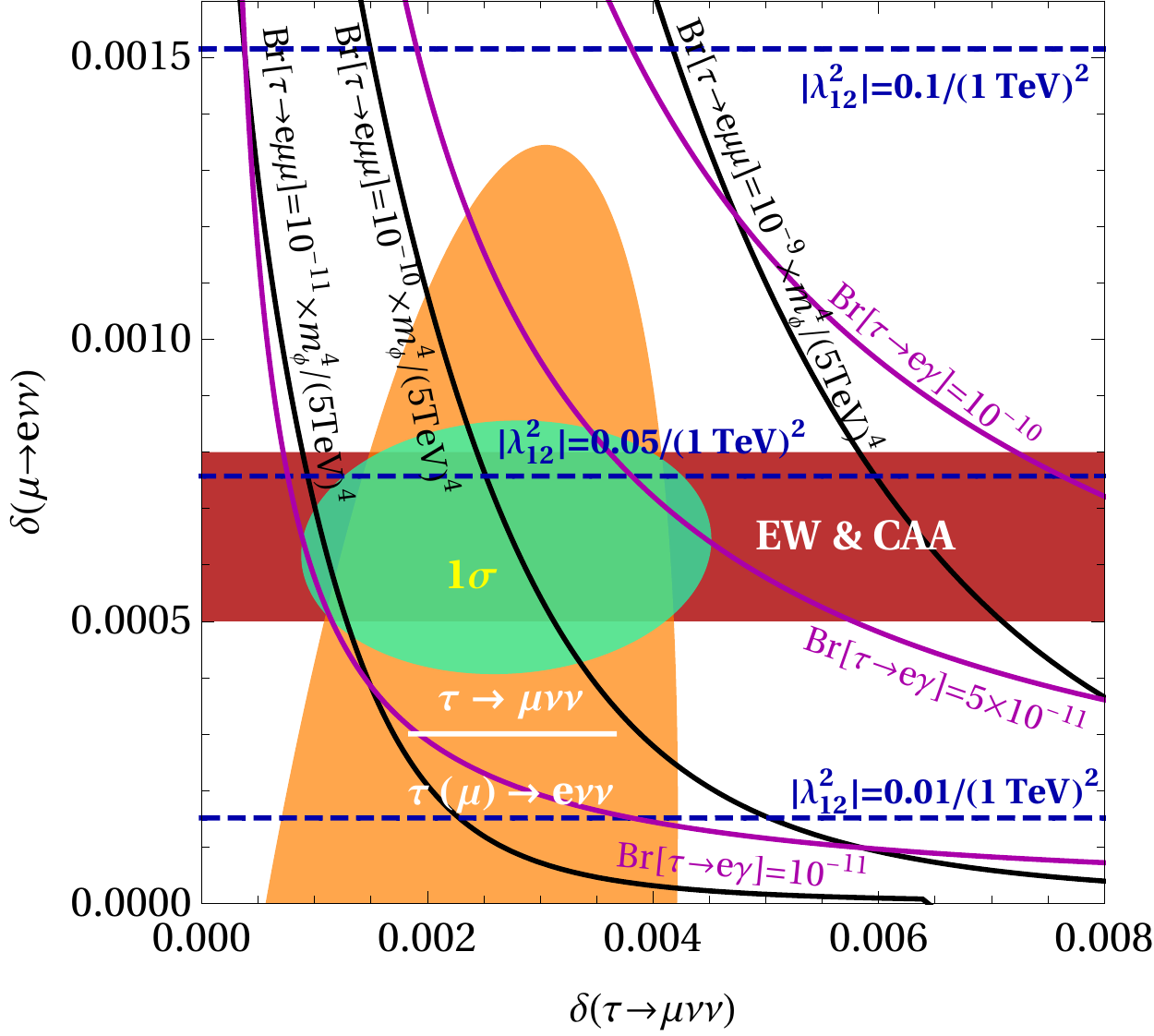}
	\caption{Regions in the $\delta (\tau\to\mu\nu\nu)$--$\delta (\mu \to e\nu\nu)$ plane preferred at the level of $1\sigma$ by electroweak data, the CAA and $\ell\to\ell'\bar \nu\nu$ (see E.q.~\eqref{eq:LFUratios}). The combined region is shown in green. The curves indicate predictions for $\tau\to e\gamma$ (magenta), $\tau\to e\mu\mu$ (black) and $|\lambda_{12}^2|/m_\phi^2$ (blue).\label{fig:deltaPlot}}
\end{figure*}

Combining the flavour bounds and collider constraints in the scenario with $\lambda_{13}=0$, which is motivated by the strong constraints from $\mu\to e\gamma$ and $\mu \to e$ conversion in nuclei, we can determine best fit regions in the $\delta (\tau\to\mu\nu\nu)$--$\delta (\mu \to e\nu\nu)$ plane. In Fig.~\ref{fig:deltaPlot} the region preferred at the level of 1$\sigma$ by electroweak data and the Cabibbo Angle Anomaly is shown in red, while the region preferred at the level of 1$\sigma$ by the ratios listed in E.q.~\eqref{eq:LFUratios} is shown in orange. The combined best fit region is indicated in green. In this region of parameter space we find ${\rm Br}(\tau\to e\bar \mu\mu)\sim 10^{-10} m_\phi^4/(5~\text{TeV})^4$, $10^{-11}\lesssim {\rm Br}(\tau \to e\gamma)\lesssim 5\times 10^{-11}$ and $|\lambda_{12}|^2\sim 0.05/(1~\text{TeV})^2$, the last of which can be probed by monophoton searches at future $e^+e^-$ colliders. 

We have thus shown that the singly charged singlet scalar, which is a very predictive simplified extension of the Standard Model, can simultaneously explain the Cabibbo Angle Anomaly, as well as the hints for lepton flavour universality violation in $\tau \to \mu \bar \nu\nu$, while predicting loop-suppressed charged lepton flavour violation.

\section*{Acknowledgments}
I would like to thank the organisers of the 55$^\text{th}$ Rencontres de Moriond and of the BSM21 conference for two successful online-conferences, and Jean-Marie Fr\`ere, as well as Andreas Crivellin, for giving me the opportunity to talk at Moriond@home.


\end{document}